\documentclass[sigconf]{acmart}

\setcopyright{acmcopyright}
\copyrightyear{2023}
\acmYear{2023}
\acmDOI{XXXXXXX.XXXXXXX}

\acmConference[P3HPC '23]{International Workshop on Performance, Portability \& Productivity in HPC}{November 12, 2023}{Denver, CO}

\usepackage{xspace}
\usepackage{listings}
\usepackage{url}
\usepackage{subcaption}

\usepackage{hyperref}

\usepackage{units}
\usepackage{amsmath}
\usepackage{mathrsfs}
\usepackage{mathtools}
\usepackage{phoenician}
\newcommand{\pp}{\textrm{PP}}
\newcommand{\ppm}{$\pp$\xspace}

\definecolor{Comments}{RGB}{0, 127, 0}
\definecolor{Keywords}{RGB}{0, 0, 255}
\definecolor{Types}{RGB}{43, 145, 175}
\newcommand{\paper}{paper\xspace}
\newcommand{\fixme}[1]{\textcolor{red}{\uppercase{#1}}}

\newcommand{\Intel}{Intel\textregistered\xspace}

\newcommand{\eg}{\textit{e.g.,}\xspace}
\newcommand{\ie}{\textit{i.e.,}\xspace}        

\copyrightyear{2023} 
\acmYear{2023} 
\setcopyright{licensedusgovmixed}\acmConference[SC-W 2023]{Workshops of The International Conference on High Performance Computing, Network, Storage, and Analysis}{November 12--17, 2023}{Denver, CO, USA}
\acmBooktitle{Workshops of The International Conference on High Performance Computing, Network, Storage, and Analysis (SC-W 2023), November 12--17, 2023, Denver, CO, USA}
\acmPrice{15.00}
\acmDOI{10.1145/3624062.3624187}
\acmISBN{979-8-4007-0785-8/23/11}

\begin{document}

\title[A Performance-Portable SYCL Implementation of CRK-HACC for Exascale]{A Performance-Portable SYCL Implementation of\\CRK-HACC for Exascale}

\author{Esteban M. Rangel}
\email{erangel@anl.gov}
\orcid{0000-0003-2330-9820}
\affiliation{%
  \institution{Argonne National Laboratory}
  \country{USA}
}

\author{S. John Pennycook}
\email{john.pennycook@intel.com}
\orcid{0000-0003-0237-3823}
\affiliation{%
  \institution{Intel Corporation}
  \country{USA}
}

\author{Adrian Pope}
\email{apope@anl.gov}
\orcid{0000-0003-2265-5262}
\affiliation{%
  \institution{Argonne National Laboratory}
  \country{USA}
}

\author{Nicholas Frontiere}
\email{nfrontiere@anl.gov}
\orcid{0009-0005-8598-4292}
\affiliation{%
  \institution{Argonne National Laboratory}
  \country{USA}
}

\author{Zhiqiang Ma}
\email{zhiqiang.ma@intel.com}
\orcid{0009-0008-1169-2205}
\affiliation{%
  \institution{Intel Corporation}
  \country{USA}
}

\author{Varsha Madananth}
\email{varsha.madananth@intel.com}
\orcid{0009-0000-3899-0589}
\affiliation{%
  \institution{Intel Corporation}
  \country{USA}
}

\begin{abstract}

The first generation of exascale systems will include a variety of machine architectures, featuring GPUs from multiple vendors. As a result, many developers are interested in adopting portable programming models to avoid maintaining multiple versions of their code. It is necessary to document experiences with such programming models to assist developers in understanding the advantages and disadvantages of different approaches.

To this end, this paper evaluates the performance portability of a SYCL implementation of a large-scale cosmology application (CRK-HACC) running on GPUs from three different vendors: AMD, Intel, and NVIDIA. We detail the process of migrating the original code from CUDA to SYCL and show that specializing kernels for specific targets can greatly improve performance portability without significantly impacting programmer productivity. The SYCL version of CRK-HACC achieves a performance portability of 0.96 with a code divergence of almost 0, demonstrating that SYCL is a viable programming model for performance-portable applications.
\end{abstract}



\settopmatter{printacmref=false}
\setcopyright{none}
\renewcommand\footnotetextcopyrightpermission[1]{}
\pagestyle{plain}

\maketitle

\section{Introduction}\label{sec:introduction}


\noindent The US Department of Energy (DOE) established the Exascale Computing Project (ECP) as a large, coordinated, multi-year effort within the DOE high performance computing (HPC) community to ensure that scientific projects and codes would be prepared for the arrival of the first exascale systems.
Now that the first exascale systems in the US are being deployed, with the Frontier system at the Oak Ridge Leadership Computing Facility (OLCF) and the Aurora system at the Argonne Leadership Computing Facility (ALCF), we can study how specific codes have prepared for their arrival.
Though all of the flagship HPC systems operated by the DOE Office of Science at the beginning of the exascale era employ GPUs, they leverage three different architectures developed by three different vendors.
The Perlmutter system at the National Energy Research Scientific Computing Center (NERSC) and OLCF's previous Summit system both use NVIDIA GPUs, while OLCF's new Frontier system uses AMD GPUs, and ALCF's Aurora system will use \Intel GPUs.


In this \paper, we study the Hardware/Hybrid Accelerated Cosmology Code (HACC)~\cite{HACC-NewAstronomy}, a code that has been developed and maintained by researchers within the DOE HPC community for roughly the past 15 years.
From its inception, HACC was designed specifically to make use of compute nodes with discrete accelerators in addition to CPUs.
HACC has run on all of the DOE Office of Science's flagship computing systems for over a decade~\cite{HACC-BGQ-OuterRim, HACC-OpenCL-Titan-QContinuum, HACC-BGQ-LastJourney, HACC-CUDA-Summit-Farpoint}, and has often been selected to take part in programs aimed at early access to those systems.
One of HACC's primary goals has been extremely high performance, and the code has been designed so that the majority of lines of code are maintained as C++ that is portable across the CPU environments, and a sub-dominant fraction of the code represents most of the total compute intensity.
In order to achieve extremely high performance, some portions of the computationally intense code have historically been adapted specifically to each architecture, either by the use of intrinsics, assembly, or by integrating sections of code in another programming model.

Some of the work in preparing the most recent version of HACC, called CRK-HACC, for exascale systems has been pursued under ECP through the ``Computing the Sky at Extreme Scales'' (ExaSky) project.
This largely focused on using CUDA on OLCF's Summit system, and developing a mechanism to maintain a HIP wrapper around the CUDA code for use on OLCF's Frontier system.
Preparing CRK-HACC for Aurora has been pursued mainly through the ALCF's Early Science Program (ESP), and has focused on developing methods and tools to maintain SYCL versions of the CUDA kernels without constantly duplicating the continuing effort that goes into developing the CUDA kernels.
SYCL is a portable programming model designed to support a wide range of hardware architectures~\cite{SYCL-2020-r7}, and now that we have a SYCL version of CRK-HACC we can target all three of the GPU architectures of interest with the same source code.
This \paper details the process of preparing a SYCL version of CRK-HACC, explores its performance portability characteristics, and discusses programmer productivity.

Specifically, the contributions of this \paper are as follows:
\begin{itemize}
    \item We detail the migration of a large, complex, C++ code from CUDA/HIP to SYCL and provide guidance for other developers interested in migrating their own codes.\smallskip
    
    \item We provide a detailed performance analysis of the migrated SYCL code on the \Intel Data Center GPU Max Series, and present a number of code optimizations which improve performance (portability). \smallskip
    
    \item We compute and compare the performance portability and code divergence of multiple versions of CRK-HACC, to explore the trade-offs between maintaining a single source code and specializing small regions of code for specific targets. \smallskip
    
    \item We provide a retrospective on the development of the SYCL version of CRK-HACC, discussing a number of challenges faced by the authors during the migration process, with advice for developers embarking on similar efforts.
\end{itemize}

\section{Related Work}\label{sec:related}

A previous study~\cite{Argonne-P3-Miniapps} investigated the performance portability of four CRK-HACC kernels at an earlier point in its development and with earlier hardware, using automatically migrated CUDA kernels and an integrated \Intel GPU. This \paper builds upon that work by examining the performance portability of the entire CRK-HACC code, running on production discrete \Intel GPUs.

A number of other studies have detailed the process of migrating from CUDA to SYCL~\cite{Migration-PerturbativeTriples, Migration-AutoDock, Migration-LegacyStencil, Migration-SmithWaterman}. This paper provides additional insight into migrating CUDA codes that are still under active development, and codes where it is desirable to be able to reference kernels by name. Additionally, by investigating the performance (portability) of the migrated code on modern platforms, we are able to identify potential improvements to the migration process that could improve the out-of-box experience for future developers.

There have also been several studies that have investigated the performance portability characteristics of SYCL~\cite{P3-Dslash, SYCL-DNN}, typically concluding that SYCL performs well relative to other programming models. However, Johnston et al.~\cite{P3-SYCL-Implementations} note that the performance portability characteristics of SYCL applications are tied to the specific features and implementations used---the evaluation in this \paper offers new insight into the current state-of-the-art in using SYCL to develop performance-portable N-body applications.

\section{Background}\label{sec:background}

\subsection{HACC}\label{sec:HACC}


\noindent The code that became the basis for the HACC project~\cite{HACC-RRU, HACC-Pope-CiSE} was originally developed as a gravity-only cosmological N-body structure formation code written for Los Alamos National Laboratory's IBM Roadrunner supercomputer, which featured IBM Cell Broadband Engine accelerators.
HACC employed force-splitting techniques where the long-range component of gravity was calculated with particle-mesh methods and a distributed-memory Fourier transform-based Poisson solver implemented in MPI.
The short-range component of gravity was calculated using direct particle-particle comparisons, and it was implemented in C with intrinsics to take advantage of the FLOPS available on the Cell accelerators.

After Roadrunner, the development of HACC's short-range gravity solver was split into two major branches for different architectures, while the long-range Poisson solver was maintained as a single code-base for all architectures.
The branch of HACC intended for multi-core and many-core CPUs used the OpenMP 3 threading model and implemented Recursive Coordinate Bisection (RCB) tree methods to reduce the total number of particle comparisons at the cost of more complicated data structures and flow control.
The very inner kernel of the RCB version of HACC's short range gravitational force could still be optimized for specific architectures, and versions were maintained for the IBM Blue Gene/Q~\cite{HACC-BGQ-SC12} and for the \Intel Advanced Vector Extensions 512 (\Intel AVX-512) instruction set used by \Intel Xeon Phi\texttrademark~processors and coprocessors.
The branch of HACC intended for accelerators and GPUs kept the direct particle-particle comparison approach but re-implemented it in OpenCL~\cite{HACC-OpenCL-Titan-SC13}.
When it became clear that several generations of GPU-accelerated DOE supercomputers were going to use NVIDIA GPUs, the OpenCL branch of HACC was ported to CUDA due to the availability of better debugging and performance profiling tools.

In anticipation of the additional FLOPS that would become available on exascale supercomputing systems, the HACC development team started a long term project to add baryonic physics in addition to gravity in order to increase the fidelity of galaxy groups and clusters within cosmological simulations.
Members of the team developed a new Conservative Reproducing Kernel (CRK) formulation~\cite{HACC-CRK-SPH} of Smoothed Particle Hydrodynamics (SPH) with the goal of resolving some discrepancies with grid-based hydrodynamic schemes while maintaining the scaling and performance of a particle-based scheme.
While the gravity-only version of HACC modeled a single species of effectively collision-less particles, CRK-HACC~\cite{HACC-CRK-HACC} uses separate particle species to model both dark matter, which only responds to gravity, and baryonic matter, which responds to additional forces.
The first version of CRK-HACC~\cite{HACC-BorgCube} modeled non-radiative hydrodynamics, which is often referred to as running in ``adiabatic'' mode.
Going beyond adiabatic mode involves additional sub-grid models for baryonic physics, including radiative cooling, star formation, and feedback from supernovae and Active Galactic Nuclei (AGN).
The bulk of the numerical intensity in CRK-HACC is in the adiabatic kernels.
The sub-grid kernels are less numerically intense themselves, but they can affect the time-stepping criteria and lead to many more calls to the adiabatic kernels in order to converge over the same span of cosmological time compared to running in adiabatic mode.
As testing higher fidelity modes of CRK-HACC became more numerically demanding, the CPU-only version of CRK-HACC became essentially unmaintained, and all development on CRK-HACC shifted to CUDA with testing on NVIDIA GPUs.
More recently, CRK-HACC has drawn from experiences porting other applications to Frontier~\cite{ReadyFrontier} to also support compilation with HIP. A simple header file with C-preprocessor macros is used to map CUDA functions to their HIP equivalents, allowing HACC to target AMD GPUs without first being migrated to the HIP API.

In some cases the sub-grid kernels also introduce considerable additional complexity.
For example, modeling AGN feedback requires periodically identifying the massive dark matter halos that can host AGN.
HACC had implemented a host-side dark matter halo finder based on a Friends-of-Friends (FOF) algorithm, and while it was sufficiently fast to be used for occasional in-situ analysis, it was too slow to be used at the frequency required for the AGN feedback kernels.
Members of the HACC team worked with developers of ArborX~\cite{ArborX}, a Kokkos~\cite{Kokkos}-based open-source library designed to provide performance portable algorithms for geometric search, to develop a version of the DBSCAN algorithm that could be used to execute the FOF algorithm efficiently using Kokkos backends, which are available for several GPU architectures.

\subsection{Performance Portability}\label{sec:pp}

\noindent This \paper measures performance portability using a metric~\cite{PP-PMBS, PP-FGCS} that is calculated as follows:
\begin{equation}
 \pp(a,p,H) =
 \begin{cases}
  \dfrac{|H|}{\sum\limits_{i \in H} \dfrac{1}{e_i(a,p)}} & \parbox[t]{.15\textwidth}{\text{if} $\forall i \in H$ \newline $e_i(a,p) \neq 0$}\\
  0                                             & \text{otherwise} \\
 \end{cases}
 \label{eq:pp}
\end{equation}

\noindent where $a$ is an application, $p$ is a specific input problem, $H$ is the set of platforms of interest, and $e_i(a, p)$ is the efficiency with which application $a$ solves problem $p$ on platform $i$.  If application $a$ does not run correctly on any platform $i$ in $H$, then the application cannot be considered portable (across the platforms in $H$) and its performance portability score is therefore 0.

Note that \ppm deliberately ignores productivity considerations, and places no restrictions on how an application's source code is structured in order to achieve portability. Applications which employ code paths specialized for specific targets are not penalized relative to applications which use a single source code.

For more detailed discussion of how to interpret \ppm values, the reader is referred to our previous publications~\cite{PP-Interpreting, PP-Revisiting}.

\subsection{Code Divergence}\label{sec:cd}

\noindent Programmer productivity is notoriously hard to measure. In this \paper, our analysis focuses solely on the impact of manual code specialization upon programmer productivity during maintenance tasks (\eg adding new features, or enabling new platforms).

We measure this impact using a version of the code divergence metric first proposed by Harrel, Kitson et al.~\cite{P3HPC-Code-Divergence},which represents the average pair-wise distance between source codes used to target different platforms. Code divergence is calculated as:

\begin{equation}\label{eq:cd}
    \mathrm{CD}(a,p,H) = \binom{|H|}{2}^{-1} \mkern-25mu \sum_{\{i, j\} \in H \times H} \mkern-18mu {d_{i, j}(a, p)}
\end{equation}

\noindent where $d_{i, j}(a, p)$ represents the distance between the source code required to solve problem $p$ using application $a$ on platforms $i$ and $j$ (from platform set $H$).

Building on our previous work~\cite{NavigatingP3}, we calculate the distance between two source codes using the Jaccard distance:

\begin{equation}\label{eq:jaccard}
   d_{i, j}(a, p) = 1 - \frac{|c_i(a, p) \cap c_j(a, p)|}{|c_i(a, p) \cup c_j(a, p)|}
\end{equation}

\noindent where $c_i$ and $c_j$ represent the set of source lines required to compile application $a$ and execute problem $p$ for platforms $i$ and $j$, respectively.

Values of this metric fall in the range $[0, 1]$: a value of $0$ means that all code is shared (\ie there is no specialization for any platform); and a value of $1$ means that no code is shared (\ie the entire code is specialized for each platform).

\subsection{Experimental Setup}\label{sec:setup}

\begin{table*}[ht!]
    \begin{tabular}{l|l|l|l|l|l}
        \textbf{System} & \textbf{CPU} & \textbf{Sockets} & \textbf{GPU} & \textbf{\# GPUs} & \textbf{FP32 Peak per GPU} \\ \hline
        Aurora
        & Intel\textregistered~Xeon\textregistered~CPU Max 9470C, 52 cores
        & 2
        & Intel\textregistered~Data Center GPU Max 1550
        & 6
        & 45.9 TFLOPS \\
        Polaris
        & AMD EPYC 7543P, 32 cores
        & 1
        & NVIDIA A100-SXM4-40GB
        & 4
        & 19.5 TFLOPS \\
        Frontier
        & AMD EPYC 7A53, 64 cores
        & 1
        & AMD Instinct MI250X
        & 4
        & 53 TFLOPS \\
    \end{tabular}
    \caption{Hardware configuration for one node of each test system.}\label{tab:systems}
\end{table*}

\subsubsection{Systems}

\noindent For this study we used three systems representing three major GPU architectures from Intel, NVIDIA, and AMD.
For \Intel GPUs, we used ALCF's Aurora\footnote{\url{https://www.alcf.anl.gov/aurora}} system.
For NVIDIA GPUs, we used ALCF's 
Polaris\footnote{\url{https://docs.alcf.anl.gov/polaris/hardware-overview/machine-overview/\#polaris-a100-gpu-information}}
production system with A100 GPUs. 
We did not have access to NVIDIA's latest H100 GPUs for this study, but would expect similar findings with respect to portability.
For AMD GPUs, we used OLCF's Frontier\footnote{\url{https://docs.olcf.ornl.gov/systems/frontier_user_guide.html\#gpus}} system.
Table~\ref{tab:systems} summarizes the most important hardware configuration information for each system.
Full configuration information, including details of the software stack, is available in Appendix~\ref{sec:reproducibility}.

\subsubsection{Problem Size}

We use a test problem designed to run on a small number of nodes and devices while still being representative of the on-node workload during a production run.
In production, CRK-HACC usually assigns one MPI rank per accelerator device, and currently requires a minimum of 8 MPI ranks to run correctly.
On Frontier, each AMD Instinct MI250X GPU consists of 2 Graphics Compute Dies (GCDs), and each GCD is presented to the user as a logically separate accelerator device; we run 8 MPI ranks on a single node of Frontier, with each rank mapped to its own GCD. Similarly on Aurora, each Intel\textregistered~Data Center GPU Max 1550 consists of 2 compute stacks that can operate independently; we also run 8 MPI ranks on a single node of Aurora, using 2 stacks on each of 4 GPUs and leaving 2 GPUs per node idle.
The situation on Polaris is somewhat different.
Although NVIDIA's Multi-Instance GPU (MIG) feature can partition GPUs into separate instances, on A100 there are 7 available instances and there is no way to evenly divide the hardware between 2 MPI ranks while also making use of all available hardware.
Our remaining options were to use 1 Polaris node with 2 MPI ranks per GPU, or to use 2 Polaris nodes with 1 MPI rank per GPU. We decided to use a single Polaris node to maintain parity in the total number of GPUs used on each system. This results in ${\sim}11\%$ lower efficiency on Polaris.

To choose a relevant problem size for 8 MPI ranks, we looked at configurations used to calculate an application Figure-of-Merit (FOM) for CRK-HACC by the ExaSky project, a part of DOE's Exascale Computing Project (ECP).
The ExaSky project used two different problem sizes to assess the CRK-HACC FOM on 8192 nodes of Frontier.
The default problem size corresponds to $2\rm{x}\ 229^3$ particles per GCD (where the $2\rm{x}$ indicates running an equal number of dark matter and baryon particles), and the stretch problem size corresponds to $2\rm{x}\ 305^3$ particles per GCD.
We chose to use a configuration with $2\rm{x}\ 512^3$ particles divided between 8 MPI ranks, which corresponds to $2\rm{x}\ 256^3$ particles per GCD on Frontier, indicating a scaled-down problem size in-between the default and stretch FOM problem sizes.
Per node, this problem size is representative of production scale simulations and corresponds to a device memory usage of ${\sim} 10$ GB per MPI rank, which should fully exercise the memory hierarchy on each architecture.
Compute intensity is sensitive to the physical mass resolution of the simulation, so we scaled down the simulation box size to $177\ \rm{Mpc/h}$ per side to maintain the same mass resolution as the Frontier FOM problems.

\subsubsection{Physics}

For this initial study we focus on the early universe near the beginning of a simulation, running five time steps from an initial redshift $z_i=200$ to a final redshift $z_f=50$.
For adiabatic hydrodynamics, the same set subset of kernels will represent the vast majority of compute intensity near the beginning of the simulation as near the end.
While the relative workload balance could shift somewhat within this subset of kernels as the matter fields become more highly clustered, optimizing this same subset of kernels will help at all epochs.
In future we plan to study the additional sub-grid kernels involved in the full hydrodynamics calculations, and for those studies we would need to profile the simulation at much later times (since many of the sub-grid kernels are relatively inactive near the beginning of the simulation).
Even with the sub-grid kernels enabled, we would still expect the majority of the compute intensity to remain in the adiabatic and gravitational kernels which are represented in this study.
By focusing on the early portion of an adiabatic simulation we can also exclude the Kokkos portions of the CRK-HACC code, since those are only needed by the AGN feedback sub-grid kernels.

\subsubsection{Timing}

We focus our analysis on the performance and portability of the code running on the accelerator devices, rather than the full application code.
The detailed breakdown of total wall clock for a production run of CRK-HACC depends on the mass resolution, the in-situ analyses performed, and time spent in IO.
But the vast majority of wall clock for the dynamical time stepping is spent in the adiabatic hydro and short-range gravitational kernels on the GPUs, and only a small fraction of time goes to host-side operations like the 3D distributed-memory FFTs for the Poisson solver.
For this study we disabled all in-situ analysis and IO, so all wall clock time after initialization can be attributed to the dynamical time stepper.

CRK-HACC has various internal timers that use \texttt{MPI\_wtime()} to bracket various operations and report timings as a simulation progresses.
These timers have proven very reliable for operations that take more than ${\sim}1ms$, and during this study we validated CRK-HACC's timers for the major GPU kernels against times reported by \texttt{rocprof} on AMD Instinct MI250X and found very good agreement.
In addition to timers for individual kernels, we also used a CRK-HACC timer that brackets all of the offloaded GPU operations during the dynamical time stepping to track the total time spent in code running on the accelerator devices.


\section{Migration to SYCL}\label{sec:migration}

SYCL is an industry standard from the Khronos Group that enables developers to target heterogeneous systems using C++.  The latest SYCL specification (SYCL 2020~\cite{SYCL-2020-r7}) allows implementations to support multiple ``backends'', and implementations are already available for multiple hardware architectures, leveraging backends including OpenCL, CUDA and HIP. 

Migrating CRK-HACC from CUDA to SYCL therefore not only enables support for \Intel GPUs, but also retains the code's ability to target NVIDIA and AMD GPUs. In this section, we provide details about the migration process, and explore the performance of the code resulting from this initial migration across different GPUs.

\subsection{SYCLomatic}\label{sec:syclomatic}

CRK-HACC contains approximately 30,000 lines of CUDA code (including comments and whitespace) spread over more than 50 files, which would have been very difficult to migrate by hand.  Instead, we used SYCLomatic~\cite{DPCT-IWOCL, SYCLomatic-IWOCL}, an open-source project which provides tooling for migrating applications from CUDA to SYCL.

When SYCLomatic encounters code that cannot be automatically migrated, or code that cannot be guaranteed to migrate safely (because of differences between the CUDA and SYCL programming models), it emits a diagnostic alerting the developer that some manual attention may be required. In the case of CRK-HACC, diagnostics were generated only for CUDA intrinsics that could be safely removed (\eg \texttt{\_\_ldg}) and for math functions with different precision guarantees (\eg \texttt{frexp}).

\subsection{Additional Tooling}

As detailed in Section~\ref{sec:HACC}, CRK-HACC supports both CUDA and HIP by providing macros and wrapper functions that encapsulate the differences between the host code for the two programming models. The SYCL equivalents of these macros and wrapper functions were straightforward to develop manually. However, the SYCL kernels produced by SYCLomatic were not compatible with the wrappers and required some adjustment.

CRK-HACC's kernel launch abstraction assumes that kernels can be referenced by name, as when launching a CUDA kernel (see \figurename~\ref{fig:launch-cuda}). At the time of writing, SYCLomatic migrates CUDA kernel functions to C++ functions, and replaces each CUDA kernel launch with a submission of an unnamed lambda (see \figurename~\ref{fig:launch-lambda}). We solve this problem by defining SYCL kernels as function objects instead (see \figurename~\ref{fig:launch-object}).

\begin{figure}
    \begin{subfigure}{0.45\textwidth}
    \begin{lstlisting}
// CUDA kernel defined as a function invoked by <<<>>>
__global__ void cuda_kernel(float* a, int b) {
  /* kernel body */
}
...
cuda_kernel<<<...>>>(a, b);
    \end{lstlisting}
    \caption{Launching a kernel with CUDA.}\label{fig:launch-cuda}
    \end{subfigure}
    \begin{subfigure}{0.45\textwidth}
    \begin{lstlisting}
// SYCL kernel defined as a function invoked by a kernel lambda
void sycl_kernel(float* a, int b, sycl::nd_item<3> it) {
  /* kernel body */
}
...
q.parallel_for(sycl::nd_range<3>(...),
               [=](sycl::nd_item<3> it) {
  sycl_kernel(a, b, item_ct1);
});
    \end{lstlisting}
    \caption{Launching a SYCL kernel lambda.}\label{fig:launch-lambda}
    \end{subfigure}
    \begin{subfigure}{0.45\textwidth}
    \begin{lstlisting}
// SYCL kernel defined as a function object invoked directly
struct SYCLKernel {

  void operator()(sycl::nd_item<3> it) {
    /* kernel body */
  }

  const float* a;
  const int b;
};
...
q.parallel_for(sycl::nd_range<3>(...), SYCLKernel(a, b));
    \end{lstlisting}
    \caption{Launching a SYCL kernel function object.}\label{fig:launch-object}
    \end{subfigure}
    \caption{A comparison of various kernel launch forms.}\label{fig:lambda-vs-functor}
\end{figure}

Function object transformation is the first stage in a short migration pipeline that performs the complete source-to-source kernel translation (\eg header substitution, replacement of SYCLomatic helper functions from the \texttt{dpct} namespace, and insertion of our own wrappers for common operations like shuffles and atomics). The functor tool itself is based on Clang's \texttt{LibTooling} and traverses the abstract syntax tree (AST) to identify the kernel functions in each compilation unit, leveraging the \texttt{Rewriter} and \texttt{SourceManager} classes to make the necessary edits. The tool generates a header file for each kernel that contains the function object's class declaration and constructor(s), but defines the call operator containing the kernel body in the original source file to maintain a similar file structure to the original code base.

Another advantage of defining kernels as function objects is that we can leverage other features of C++ (\eg templates, class inheritance). Common functionality can be encapsulated in a base class shared by all kernels, and certain properties of kernels (\eg the required amount of work-group local memory) can be encoded in the kernel type using compile-time constant expressions.

\subsection{Additional Fixes}

One key difference between CUDA and SYCL is the difference between warps and sub-groups. In CUDA, thread blocks are divided into warps containing 32 threads; every kernel uses the same warp size, and all existing NVIDIA hardware has a warp size of 32. In SYCL, work-groups are divided into sub-groups containing an implementation-defined number of work-items, which may vary across different kernels and devices. If a SYCL kernel requires a specific sub-group size for correctness, then it must be decorated with the \texttt{[[sycl::reqd\_sub\_group\_size(S)]]} attribute. Different devices support different values of \texttt{S}: AMD GPUs support sub-group sizes of 32 and 64, \Intel GPUs support sub-group sizes of 16 and 32, and NVIDIA GPUs support a single sub-group size of 32.

During the initial migration process, we forced all SYCL kernels to be compiled with a sub-group size of 32, in order to minimize the differences between the two programming models and simplify debugging. However, we set the sub-group size via a macro to simplify recompiling and testing with different sub-group sizes. 

\subsection{Initial Results}

\begin{figure}
    \includegraphics[width=0.45\textwidth]{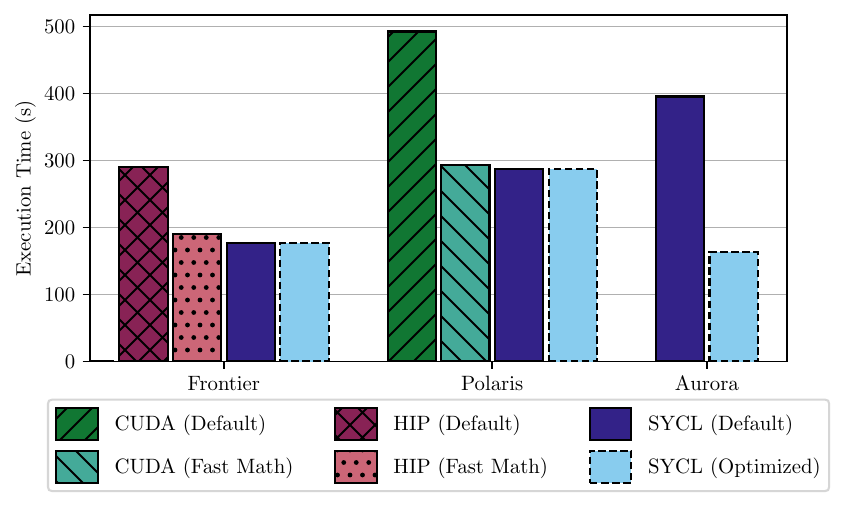}
    \caption{The initial performance of the migrated SYCL code compared to CUDA, HIP, and our optimized SYCL code.}\label{fig:initial}
\end{figure}

\figurename~\ref{fig:initial} shows the initial performance of the migrated SYCL code, measured as the execution time of all GPU kernels. On Frontier, the HIP code uses a wavefront size of 64, and the SYCL code uses a sub-group size of 64. Similarly, on Polaris the CUDA code uses a warp size of 32, and the SYCL code uses a sub-group size of 32. On Aurora, the SYCL code uses a sub-group size of 32.

Surprisingly, the results of our initial migration show SYCL significantly outperforming both CUDA on Polaris and HIP on Frontier. By comparing the intermediate PTX representations from CUDA and SYCL on Polaris, we identified approximate math functions coming from the SYCL code generation to be the root cause for this discrepancy. There are differences between the default behaviors of the compilers used: the oneAPI DPC++ compiler defaults to fast math, whereas \texttt{nvcc} and \texttt{hipcc} do not. Recompiling the CUDA and HIP codes with fast math flags closes this gap.

Even when using fast math, the SYCL code is slightly faster than both CUDA and HIP. However, the breakdown of individual kernel timings (not shown) reveals that some kernels are slightly faster and some are slightly slower. We believe this is due to the different compilers choosing different optimizations, with possibly different heuristics guiding how and when they should be applied~\cite{HIPSYCL-V100}.

Also surprising is that the initial results on Aurora are not in line with the differences in theoretical peak across the machines. As shown in Table~\ref{tab:systems}, the theoretical peaks for the GPUs on Aurora and Frontier are very similar, and so we should expect similar levels of performance. Section~\ref{sec:optimization} explains the necessary steps to improve performance on Aurora, and thus the overall performance portability of CRK-HACC. With these optimizations (including using a sub-group size of 16), performance improves by 2.4$\times$, as shown by the final bar in \figurename~\ref{fig:initial}.

\section{Optimization}\label{sec:optimization}

The optimizations employed in this section were primarily identified by inspecting the migrated SYCL code and associated assembly. We focus on the five key hot-spots in CRK-HACC that account for over 85\% of the solver's offloaded execution time. Namely, these are the following kernels: \emph{Geometry}, which measures the volumes of gas particles; \emph{Corrections}, which computes the reproducing kernel coefficients of the higher order smoothed particle hydrodynamics (SPH) solver; \emph{Extras}, which evaluates the density and state gradients; \emph{Acceleration}, which calculates the momentum derivative; and \emph{Energy}, which solves the derivative of the internal energy. 

\subsection{Hardware-Agnostic Optimizations}

Inspecting the migrated SYCL code closely highlighted several optimization opportunities that we consider hardware-agnostic. Each of the code changes discussed in this section uses a feature that is not available in CUDA, and which gives SYCL implementations additional information to inform optimizations. Importantly, all these features are part of the SYCL 2020 standard, and so are guaranteed to be supported across all architectures of interest.

First, we were able to replace several uses of warp shuffle intrinsics with higher-level algorithms from SYCL's group algorithms library.  For example, shuffles where all work-items read from the same source index can be replaced by \texttt{group\_broadcast}, and shuffle networks used to perform summations can be replaced by\linebreak \texttt{reduce\_over\_group}. These higher-level algorithms convey more information about the communication pattern to the compiler, enabling the compiler to perform more aggressive optimizations.

Second, we were able to swap several uses of built-in math functions (\eg \texttt{sycl::pow}) for native equivalents with reduced precision requirements and domain restrictions (\eg \texttt{sycl::native::powr}). Again, these changes enable more aggressive optimizations (including the use of fast approximation instructions, where available).

Third, we were able to simplify some index calculations using SYCL built-ins. In CUDA, it is common to calculate warp-related indices using integer division and modulo operations. In SYCL, the sub-group index and sub-group size can be queried directly using member functions of the \texttt{sycl::sub\_group} class. By using the built-ins, we can potentially avoid some expensive integer operations (on hardware where the sub-group indices are known, or stored in special registers), and also improve the ability of compilers to reason about data access patterns and data uniformity.

Finally, we were able to use the modern \texttt{sycl::atomic\_ref} class from SYCL 2020 to simplify and improve several usages of atomics. CUDA's \texttt{atomicMin} and \texttt{atomicMax} functions are limited to integer types, presumably reflecting the capabilities of NVIDIA GPUs. SYCL, on the other hand, exposes \texttt{fetch\_min} and \texttt{fetch\_max} for floating-point types across all hardware. If a SYCL program is compiled for a device that does not provide native support for these operations (like NVIDIA GPUs), then the operation is emulated using an atomic compare-and-swap instruction.

\subsection{Relieving Register Pressure}

Some of the kernels in CRK-HACC are very large, leading to a significant number of register spills. \Intel GPUs offer two orthogonal controls to alleviate register pressure: 1) adjusting the size of the register file; and 2) adjusting the sub-group size.

By default, every thread on the Intel\textregistered~Data Center GPU Max 1550 has 128 registers. This can be increased to 256 registers by using compiler flags or kernel attributes. However, increasing the number of registers reduces the number of threads per EU from 8 to 4, limiting achievable occupancy to 50\%. The performance impact of this option is kernel-specific, and depends on the number of threads required to hide latency.

Since CRK-HACC does not make any assumptions about sub-group size, we can also reduce the sub-group size from 32 to 16.  This reduces the number of work-items per thread, effectively doubling the number of registers available to each work-item.

Taken together, the combination of these two options can provide a 4x increase in the number of available registers per work-item. For CRK-HACC, we found that the best combination of register file size and sub-group size varied across different kernels, depending on their complexity. Almost all results in this paper all use 256 registers and a sub-group size of 32, since this gave the best performance on average; exploring the tuning of these parameters for individual kernels is left to future work.

\subsection{Optimizing Communication}\label{sec:communication}

\begin{figure}
    \includegraphics[width=0.45\textwidth]{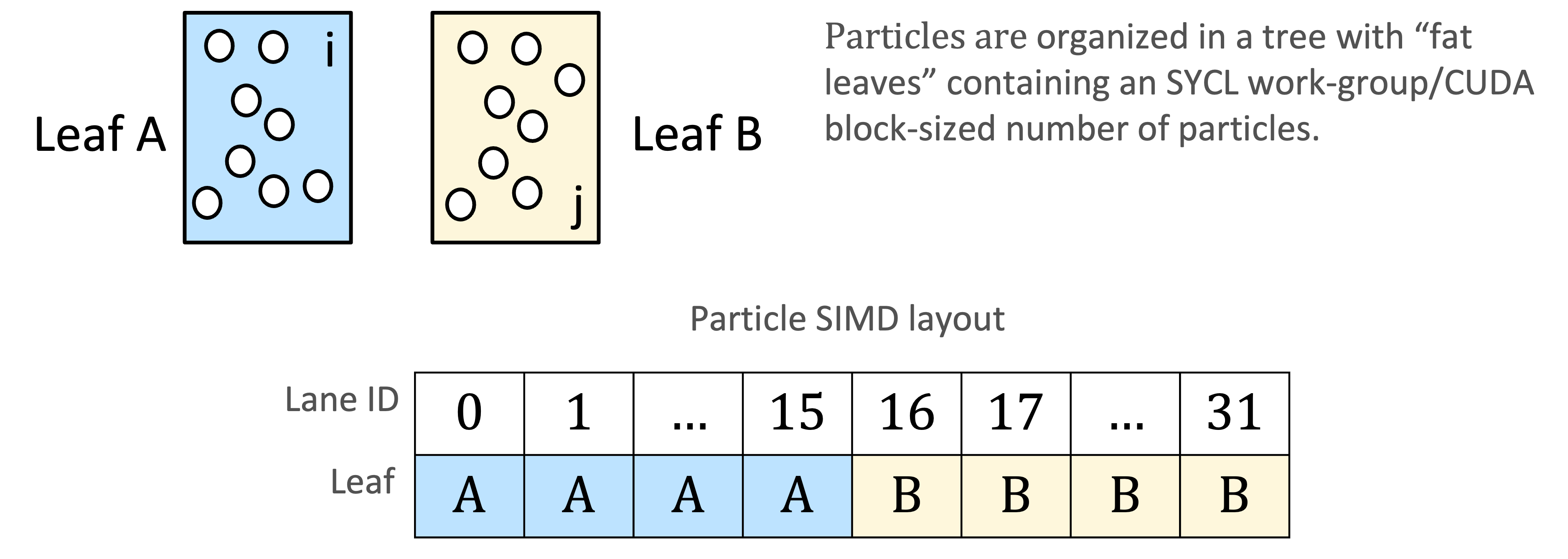}
    \caption{The SIMD lane data layout of the ``half-warp'' algorithm. Lanes [0-15] load and update particles from leaf A, while lanes [16-31] operate on particles from leaf B.}\label{fig:half-warp-layout}
\end{figure}

\noindent The CUDA version of CRK-HACC was previously optimized to alleviate register pressure by splitting the input parameters required for certain computations across two (logically defined) types of CUDA threads. This ``half-warp'' algorithm effectively specializes threads depending on their position in the warp: threads in the lower half load particles from one leaf, $i$, while threads in the upper half load particles from another leaf, $j$ (see \figurename~\ref{fig:half-warp-layout}). When a thread in the lower half evaluates the interaction $(i, j)$, another thread in the upper half evaluates the interaction $(j, i)$. The pair-wise symmetry is critically important for the correctness of the algorithm.

\begin{figure}
    \includegraphics[width=0.45\textwidth]{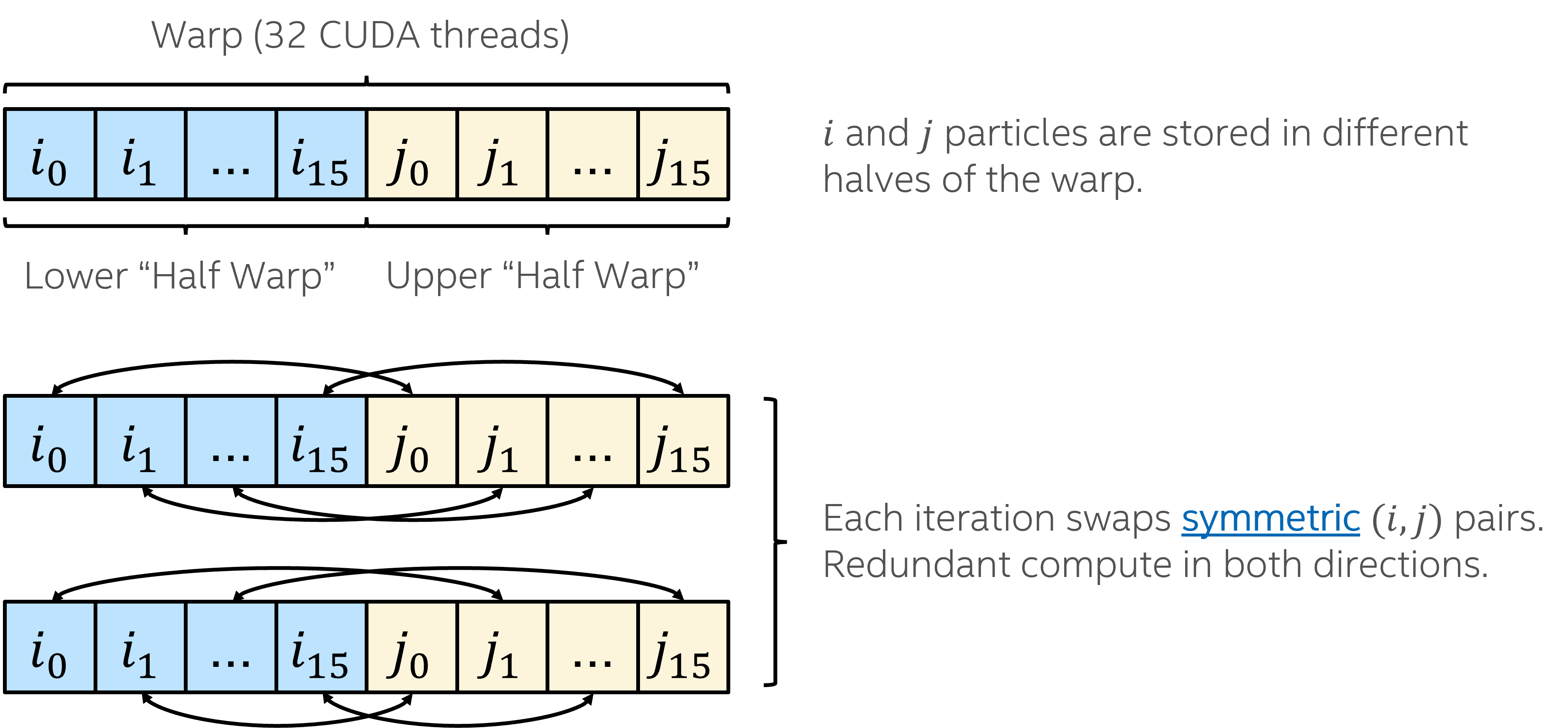}
    \caption{The communication pattern of the ``half-warp'' algorithm for interacting particles from leaves A and B within the same warp. This represents one of the total $(|Leaf_A| \times |Leaf_B| / warp\_size)$ instances required.} \label{fig:half-warp-shuffle}
\end{figure}

\begin{figure}
    \begin{lstlisting}
...
shl (16|M0)  r24.0<1>:uw  r82.0<2;1,0>:uw  0x2:uw   
add (16|M0)  a0.0<1>:uw   r24.0<1;1,0>:uw  0x640:uw 
mov (16|M0)  r2.0<1>:ud   r[a0.0]<1,0>:ud
...
    \end{lstlisting}
    \caption{Cross-lane communication implemented using indirect register access. Elements are gathered from the registers specified in \texttt{a0} and written into \texttt{r2}.}\label{fig:indirect-asm}
\end{figure}

\figurename~\ref{fig:half-warp-shuffle} shows how this communication can be implemented using an XOR-based shuffle pattern. In CUDA, this pattern is implemented via calls to the \texttt{\_\_shfl} intrinsic, and the migrated SYCL code uses \texttt{sycl::select\_from\_group}. A snippet of the assembly generated by this function on the Intel\textregistered~Data Center GPU Max 1550 is shown in Figure~\ref{fig:indirect-asm}.  The syntax \texttt{r[a0.0]} denotes an \emph{indirect register access}, whereby a vector of indices in register \texttt{a0.0} is used to gather from locations in the register file. This form of register access is relatively slow, requiring 1 cycle per element, and is the main source of inefficiency in our migrated SYCL kernels. We identified several alternative methods for optimizing this communication pattern: some are generic and straightforward to apply automatically, while others require more invasive and manual code changes.

\subsubsection{Using Local Memory}

By reserving a small amount of work-group local memory per sub-group, it is possible to swap out the use of \texttt{sycl::select\_from\_group} for a function that behaves identically, but which communicates via work-group local memory instead of via registers. Each work-item in a sub-group simply writes a value to work-group local memory, waits on a sub-group barrier, then reads out a value written by another work-item.

The implementation of this function is very straightforward, and requires only minimal changes to CRK-HACC's kernels---swapping between \texttt{sycl::select\_from\_group} and using work-group local memory is a one line change that can be implemented via a macro. Because this change is so simple, it is conceivable that future SYCL compilers could directly map usage of \texttt{sycl::select\_from\_group} to work-group local memory on the Intel\textregistered~Data Center GPU Max 1550 and thereby improve the out-of-box performance of migrated SYCL codes.  We are currently evaluating this possibility, and exploring the applicability of this optimization to other codes.

The allocation of work-group local memory is encapsulated in CRK-HACC's kernel launch abstraction wrapper by passing a \linebreak~\texttt{sycl::local\_accessor<char>} argument to all kernels. 
The number of bytes allocated is determined by the size of the largest data object to be exchanged between work-items, multiplied by the work-group size. 
The \texttt{local\_accessor} is passed as an argument to the kernel function object's constructor and used to initialize its base class. 
The communication pattern is implemented by a function defined as a method in this base class, which can be reused by all kernels. 
The function is templated (on the type being exchanged) and simply calculates the offset of the sub-groups's location in local memory relative to the work-group. 
With a pointer to this relative offset, it is trivial to implement any exchange pattern, since the memory reserved for each sub-group is guaranteed not to overlap.

\begin{figure}
    \begin{lstlisting}
...
add (16|M0)  r24.0<1>:f  r68.0<1;1,0>:f  -r14.0<0;1,0>:f 
add (16|M0)  r26.0<1>:f  r68.0<1;1,0>:f  -r14.1<0;1,0>:f
add (16|M0)  r30.0<1>:f  r68.0<1;1,0>:f  -r14.2<0;1,0>:f
...
    \end{lstlisting}
    \caption{Three broadcasts implemented using register regioning. Scalar elements are broadcast from lane 0, 1 and 2 of \texttt{r14}, subtracted from \texttt{r68}, and written into \texttt{r24}, \texttt{r26} and \texttt{r30}.}\label{fig:broadcast-asm}
\end{figure}

\subsubsection{Using Broadcasts}

On the Intel\textregistered~Data Center GPU Max 1550, only calls to \texttt{sycl::select\_from\_group} with unknown communication patterns are compiled to indirect register access.  If sufficient information is known about the communication pattern at compile-time, the compiler can generate a different instruction sequence employing \emph{register regioning}. For example, as shown in \figurename~\ref{fig:broadcast-asm}, a broadcast from a known index can be implemented by referencing a single register lane as part of an instruction---\texttt{rX.Y} refers to a scalar value stored in lane \texttt{Y} of register \texttt{X}. Such register regioning is very fast, adding negligible overhead to instruction execution.

Unfortunately, rewriting HACC's kernels to employ broadcasts instead of shuffles is not simple, and must be performed manually.  Work-items must now load data from two particles (\ie both $i$ and $j$), and in some cases must redundantly compute intermediate values that could previously be communicated between work-items. This can increase register pressure further, but has the potential to increase computational intensity (and efficiency). Due to this increased register pressure, we use a sub-group size of 16 for the broadcast kernels on \Intel GPUs. Restructuring the loops to use broadcasts also allows us to generate fewer atomic instructions.

\begin{figure}
    \includegraphics[width=0.45\textwidth]{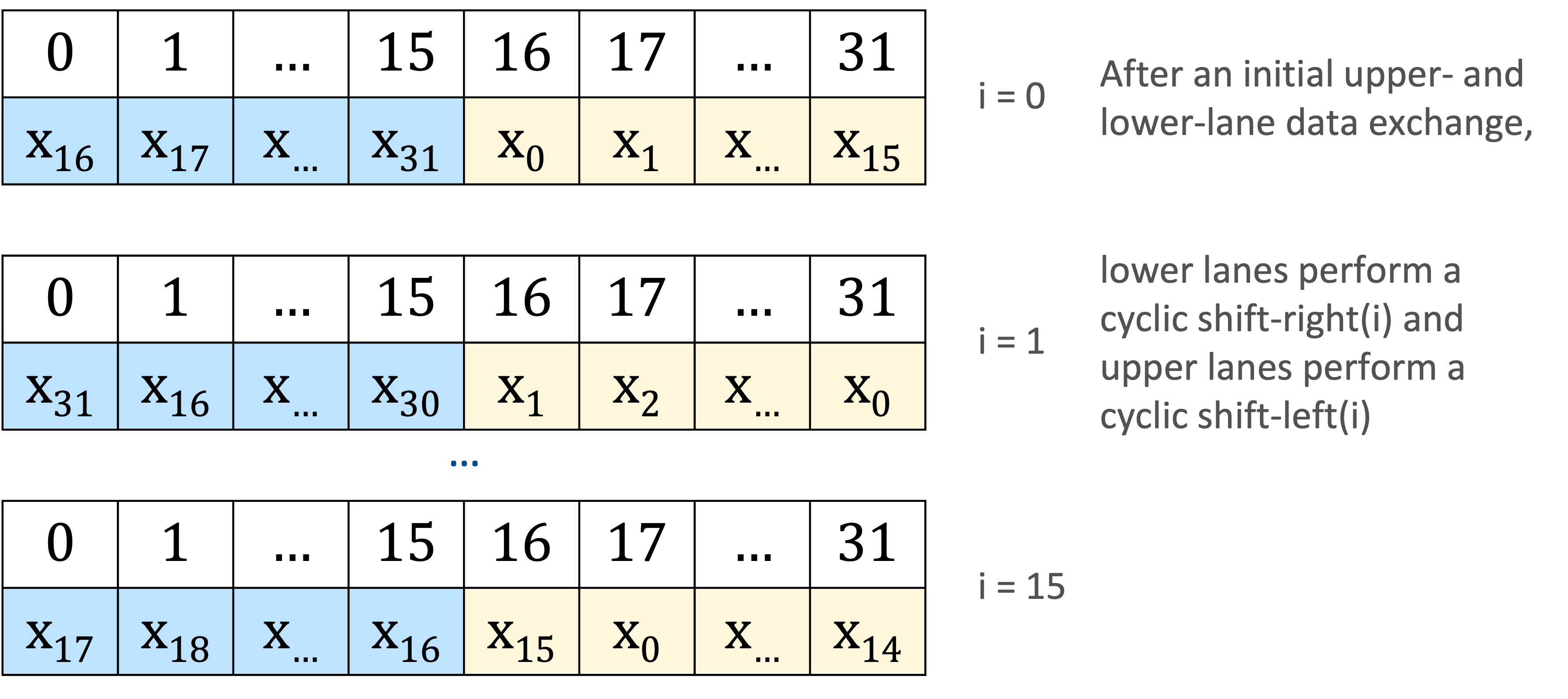}
    \caption{The specialized butterfly-shuffle communication pattern, which provides the same pair-wise symmetry property of the XOR-based pattern.}\label{fig:butterfly-shuffle}
\end{figure}

\subsubsection{Using Optimized Instruction Sequences}
It is possible to optimize the shuffle operation in registers for a slightly different exchange pattern that still preserves the critical pair-wise symmetry required by the ``half-warp'' algorithm. In \figurename~\ref{fig:butterfly-shuffle} we illustrate this specialized butterfly-shuffle, where (conceptually) after an initial upper- and lower-lane group data exchange, a cyclic inward shift is performed by the corresponding lane groups. 

\begin{figure}
    \includegraphics[width=0.45\textwidth]{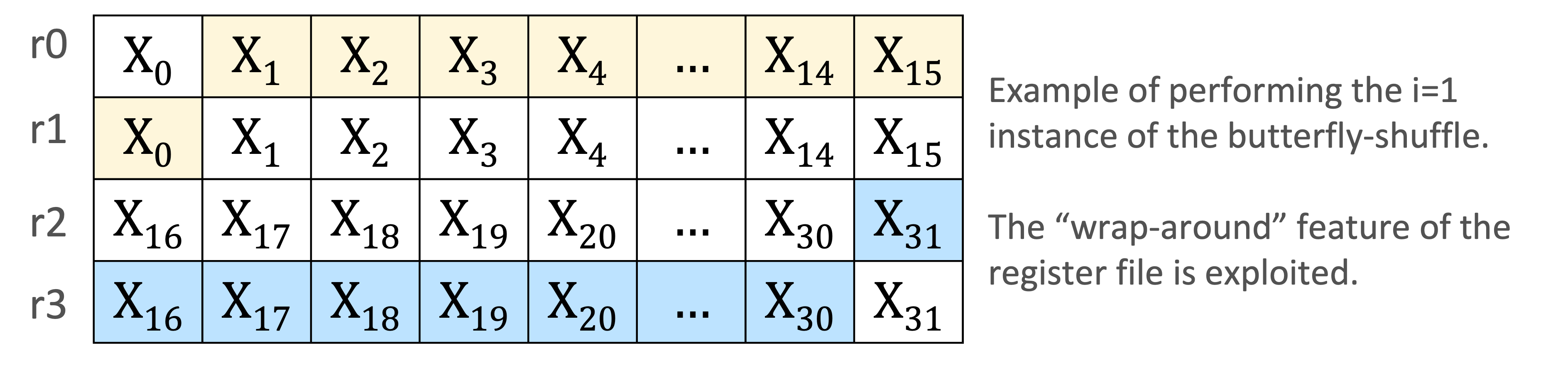}
    \caption{Register view of the specialized butterfly-shuffle.
}\label{fig:butterfly-register-view}
\end{figure}

Although the difference in this communication pattern seems subtle, it has a substantial advantage over the XOR-based pattern: namely, it can be optimized to just four \texttt{mov} instructions if the shuffle parameters are known at compile time. In \figurename~\ref{fig:butterfly-register-view} we show the register view of the second instance of the specialized butterfly-shuffle from \figurename~\ref{fig:butterfly-shuffle}. Populating the \texttt{r0} and \texttt{r1} registers is done with a single \texttt{mov} instruction, and similarly for \texttt{r2} and \texttt{r3}. The data duplication allows for the left and right shifts to be performed with register regioning, each taking one instruction.

\subsection{Optimization Results}

\begin{figure}
    \includegraphics[width=0.45\textwidth]{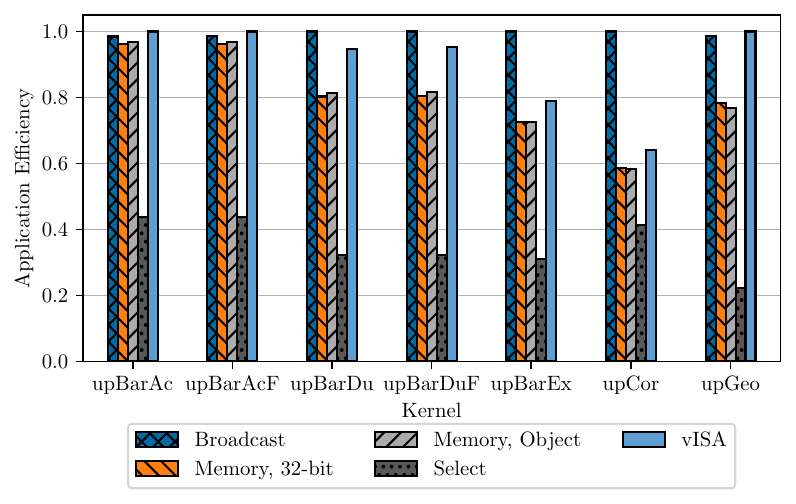}
    \caption{Application efficiency of SYCL variants on Aurora.}\label{fig:optimized-aurora}
\end{figure}

\begin{figure}
    \includegraphics[width=0.45\textwidth]{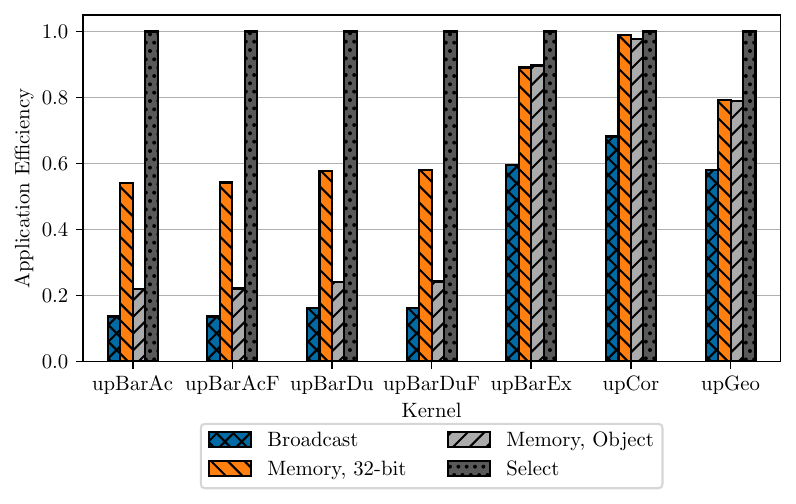}
    \caption{Application efficiency of SYCL variants on Polaris.}\label{fig:optimized-polaris}
\end{figure}

\begin{figure}
    \includegraphics[width=0.45\textwidth]{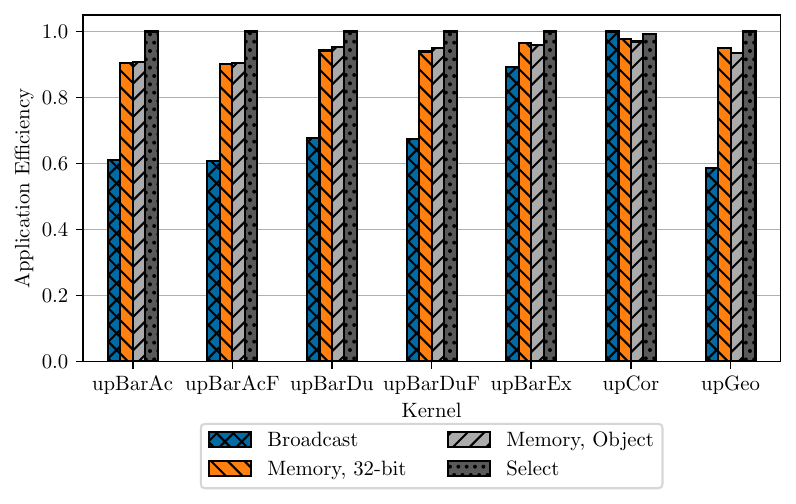}
    \caption{Application efficiency of SYCL variants on Frontier.}\label{fig:optimized-frontier}
\end{figure}

\figurename{s}~\ref{fig:optimized-aurora}, \ref{fig:optimized-polaris} and \ref{fig:optimized-frontier} show the performance of multiple SYCL variants running on Aurora, Polaris, and Frontier, respectively. Each variant corresponds to one of the optimizations described in Section~\ref{sec:communication}: \emph{Select} uses \texttt{sycl::select\_from\_group}; \emph{Memory, 32-bit} communicates via work-group local memory, exchanging each 32-bit component of composite types separately; \emph{Memory, Object} communicates via a larger work-group local memory region, exchanging composite types in one read/write; \emph{Broadcast} uses the kernels restructured for broadcasts; and \emph{vISA} uses inline vISA assembly to implement the specialized butterfly-shuffle.

To simplify comparison across graphs and kernels, performance is rendered here as application efficiency; performance results for each variant are normalized relative to the best variant on the same hardware. Some of CRK-HACC's kernels are called more than once in a single timestep, and are therefore associated with more than one wall clock timer in these graphs.  \texttt{upGeo} corresponds to geometry; \texttt{upCor} corresponds to corrections; \texttt{upBarEx} corresponds to extras; \texttt{upBarAc} and \texttt{upBarAcF} correspond to acceleration; and \texttt{upBarDu} and \texttt{upBarDuF} correspond to energy.

On Aurora, \texttt{select\_from\_group} always has the worst performance, but there is no single variant that consistently delivers the best performance. The kernels which benefit most from the broadcast optimization are those with a large number of atomic updates. Selecting the best variant for a kernel can improve performance by 2--5x, which explains the results shown in \figurename~\ref{fig:initial}; using one of the variants more suited to the architecture of \Intel GPUs delivers performance more in line with peak performance (and closes the gap between Aurora and Frontier). 

On Polaris, \texttt{select\_from\_group} always has the best performance, but the performance of the other variants remains inconsistent. The impact of choosing the ``wrong'' variant is also more pronounced, with the broadcast implementation being almost 10x slower in some cases. These slowdowns are caused by a large number of register spills in the broadcast implementation, which highlights the differences between the register files of different GPUs. The memory variants perform worst on the register heavy energy and acceleration kernels because of the trade-off between shared memory and L1 cache on NVIDIA GPUs.

On Frontier, \texttt{select\_from\_group} always has the best performance, but the ranking of the other variants is more consistent. Using work-group local memory is almost always the second-best variant (with one exception), and the broadcast variant typically has an application efficiency of $\approx 0.6$. We surmise that the MI250X is less sensitive to variant selection because it is architecturally similar to both \Intel GPUs and NVIDIA GPUs: it uses a SIMD architecture (like \Intel GPUs), but also features dedicated instructions for cross-lane communication (like NVIDIA GPUs).

That we cannot identify a single variant that delivers the best performance across all architectures and kernels highlights the difficulty of writing a single-source application that achieves high performance portability across diverse architectures.  Even though all three architectures here are GPUs, running the same code, compiled with similar compilers, they still exhibit very different affinities for different variants of the same kernel.

\section{Performance, Portability \& Productivity Analysis}\label{sec:p3}

The results in Section~\ref{sec:optimization} highlight a clear need to identify which kernel variant(s) to maintain in order to achieve an acceptable balance between application performance portability and programmer productivity. Using the metrics introduced in Sections~\ref{sec:pp} and~\ref{sec:cd} (calculated using the Performance, Portability and Productivity Analysis Library~\cite{p3-analysis-library} and Code Base Investigator~\cite{code-base-investigator}), we can quantitatively evaluate the trade-offs between several options, including: maintaining a single variant, to minimize developer effort at the expense of performance; maintaining all variants, to maximize performance portability at the expense of programmer productivity; and maintaining a subset of variants that achieve high performance portability with minimal code divergence.

\subsection{Performance Portability Analysis}\label{sec:pp-analysis}

\begin{figure}
    \includegraphics[width=0.45\textwidth]{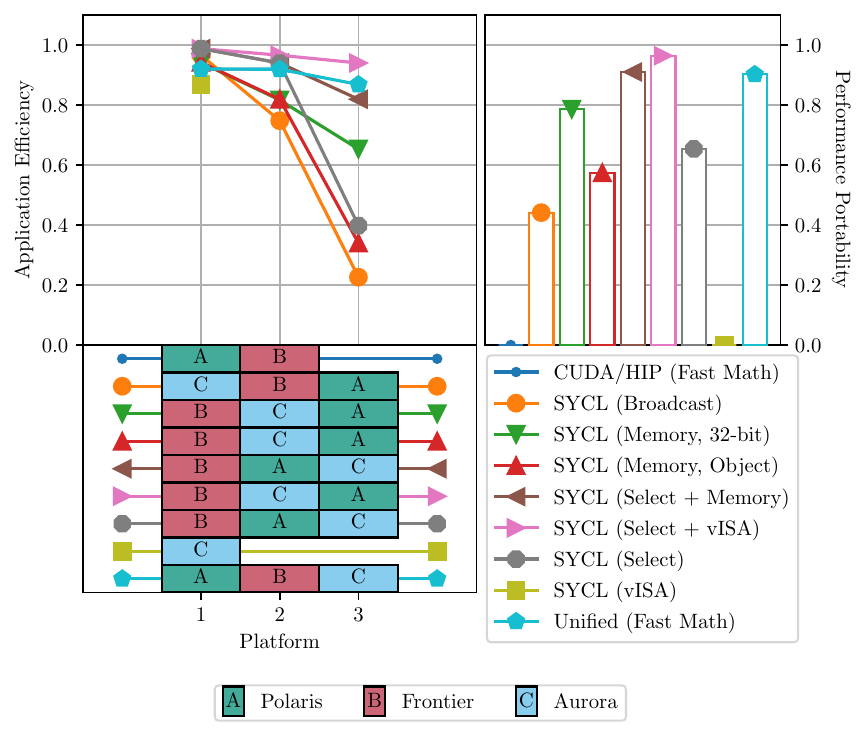}
    \caption{A cascade plot showing the application efficiency and performance portability of CRK-HACC variants.}\label{fig:pp-app}
\end{figure}

The cascade plot~\cite{PP-Interpreting} in \figurename~\ref{fig:pp-app} shows the performance portability of several different implementations of CRK-HACC. In addition to results from the CUDA, HIP, and SYCL variants discussed in previous sections, we also include results from three configurations that stitch together multiple solutions: 1) SYCL (Select + Memory), which uses \texttt{sycl::select\_from\_group} on Polaris and Frontier, but work-group local memory on Aurora; 2) SYCL (Select + vISA), which uses \texttt{sycl::select\_from\_group} on Polaris and Frontier, but inline vISA on Aurora; and 3) Unified, which uses CUDA/HIP on Polaris and Frontier, but SYCL on Aurora. In all cases, application efficiency is calculated relative to a hypothetical application that is able to use the best version of each kernel on every platform, irrespective of source language or compiler.

Two of the configurations in \figurename~\ref{fig:pp-app} (specifically, CUDA/HIP and inline vISA) have a performance portability of zero, reflecting that they do not support all platforms of interest. The CUDA/HIP code does not currently support Aurora, but there are efforts underway to enable CUDA/HIP applications to run on OpenCL and Level Zero~\cite{HIPCL, HIPLZ} in the future. The inline vISA code does not support Polaris or Frontier, and this is unlikely to change.

All other configurations run on all three platforms and thus have a non-zero performance portability. However, the performance portability of different SYCL variants is mixed, ranging from 0.44 (using broadcasts on all platforms) to 0.79 (using work-group local memory on all platforms). The latter number is respectable, but mixing and matching SYCL variants results in significantly higher performance portability; the specialized code using work-group local memory achieves a performance portability of 0.91, while the specialized code using vISA achieves a performance portability of 0.96, both of which are higher than the performance portability of 0.90 achieved from mixing CUDA, HIP and SYCL.

\subsection{Productivity Analysis}\label{sec:prod-analysis}

\begin{figure}
    \includegraphics[width=0.45\textwidth]{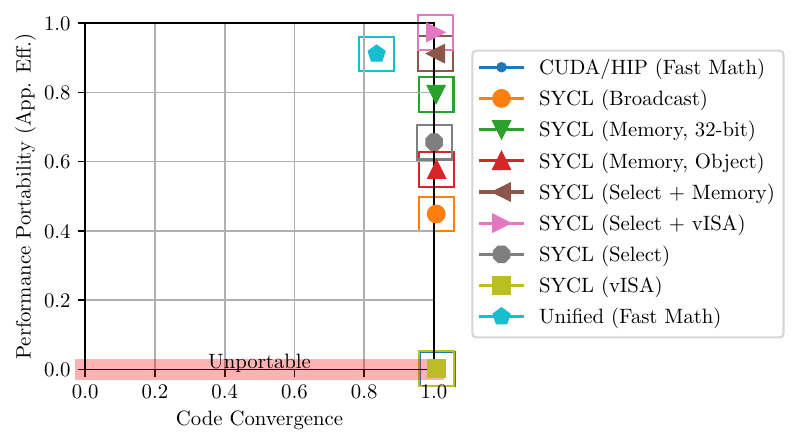}
    \caption{A navigation chart showing the performance portability and code convergence of CRK-HACC variants.}\label{fig:cd-app}
\end{figure}

The navigation chart~\cite{NavigatingP3} in \figurename~\ref{fig:cd-app} plots performance portability against code convergence (\ie $1 - \textrm{code divergence}$) for each different implementation of CRK-HACC. The best and specialized SYCL variants achieve a code convergence of almost 1.0, which highlights the fine-grained nature of our specializations (the select and work-group local memory variants differ by only 19 lines of code, and the inline assembly variant requires only an additional 226 lines of code). The only configuration with a significantly lower code convergence (0.83) is the unified variant, because it has two versions of every kernel written in both CUDA and SYCL.


\begin{table}
    \small
    \begin{tabular}{l|r|r}
        \textbf{Implementations} & \textbf{\# SLOC} & \textbf{\% SLOC} \\ \hline
        vISA & 226 & 0.27 \\
        Broadcast & 1,511 & 1.77 \\
        \textit{SYCL ($-$Broadcast)} & 1,470 & 1.73 \\
        \textit{SYCL} & 11,292 & 13.26 \\
        \hline
        HIP & 116 & 0.14 \\
		CUDA & 1,096 & 1.29 \\
        \textit{HIP and CUDA} & 6,806 & 7.99 \\
        \hline
        \textit{All} & 43,862 & 51.49 \\
        \textit{Unused} & 18,721 & 21.98 \\
        \hline \hline
        \textbf{Total} & 85,179 & 100 \\
    \end{tabular}
    \caption{Breakdown of lines of code across CRK-HACC variants. Sets containing fewer than 50 SLOC are not shown.}\label{tab:sloc}
\end{table}

Since code divergence and convergence are normalized to the size of the codebase, we also present a breakdown of raw source lines of code (SLOC)---excluding whitespace and comments---in Table~\ref{tab:sloc}. Looking at the SYCL variants, we see that the broadcast implementation is almost completely separate from the other implementations, due to the restructuring described in Section~\ref{sec:optimization}. Unexpectedly, SYCL also uses almost 1.7x as many lines as CUDA/HIP. $\approx$6,000 lines of SYCL can be attributed to the kernel function object definitions (see Section~\ref{sec:migration}), which place one kernel argument on each line and artificially inflate the line count.  It is also worth noting that these lines could be removed completely if we were not concerned with maintaining compatibility with CRK-HACC's launch wrappers. The remainder of the SYCL code (the kernels themselves) is more similar in size to the CUDA code.

Approximately 700 lines of code attributed to CUDA are a side effect of Code Base Investigator's design. Although these lines are compiled and guarded by a \texttt{\_\_CUDACC\_\_} macro, they are never actually \emph{executed}.  This is distinct from lines of code marked ``Unused'', which are not compiled at all and which correspond to features of CRK-HACC that are disabled in ``adiabatic'' mode (see Section~\ref{sec:background}).  ``Unused'' lines are excluded when calculating divergence.

\section{Retrospective and Discussion}\label{sec:retrospective}

This work is the result of a long collaboration and co-design effort between Argonne and Intel, investigating the performance of CRK-HACC kernels across several generations of hardware and multiple compiler versions. In this section, we share some of the lessons we've learned as a result of this collaboration, which may prove useful for other developers embarking on similar efforts.

\subsection{Code Divergence vs Specialization}

One of the challenges faced during this work arose due to the CUDA version of CRK-HACC being under active development. Any significant changes to the CUDA kernels had to be mirrored in the SYCL kernels---either by applying the change in both places, or by repeating the SYCL migration process with the updated CUDA code. Similarly, any optimization opportunities identified during or after the SYCL migration had to be manually backported to the CUDA version. This highlights the danger of code divergence, and its impact on programmer productivity: any duplication of logic in the code also duplicates the cost of code maintenance.


The results in Section~\ref{sec:p3} suggest that a high code divergence may not be necessary to meet HACC's performance portability goals.
The SYCL version of the code provides high levels of performance across all the platforms tested here, even when using a single version of the kernels. 
By employing specialized kernels for the top hotspots, it is possible to achieve even higher levels of performance portability while still maintaining a low code divergence. 
Provided that the SYCL programming model is available on the target HPC systems, this makes a compelling case for using SYCL as the base programming model for a performance portable application.

\subsection{Experimenting with Standalone Kernels}

To facilitate rapid prototyping and analysis, we extracted CRK-HACC's biggest hotspots into standalone applications driven by checkpoint files. The ability to quickly recompile and run one kernel at a time significantly improved our ability to experiment with potential optimizations, including changes that would be difficult to evaluate in the context of the whole code (\eg changes to data structures). Working with these standalone kernels helped us to establish an upper bound for achievable performance, and ultimately drove us to develop each of the SYCL variants outlined in Section~\ref{sec:optimization}.

\subsection{CPU Support}

Early versions of CRK-HACC had CPU code for the first few kernels, but after the CUDA version became the reference for development the CPU kernels became stale, and as new kernels were added they were only implemented in CUDA.
Before migrating the CUDA code to SYCL, there had been no attempts to run a modern version of CRK-HACC on CPUs for several years.

Although we do not explore CPU performance in this \paper (since GPU performance is our current priority), the SYCL code has been tested for correctness on CPUs using an OpenCL backend.
We expect that some additional tuning for CPUs would be required to achieve high levels of performance portability---primarily due to the way the code uses atomics---but it is interesting to note that the SYCL code is the only modern version of CRK-HACC that we have been able to run on CPUs. Portability to CPUs was not originally a goal, but we now plan to explore this further in future work.

\section{Conclusion}


\noindent We have described our process for creating a SYCL version of CRK-HACC starting from CUDA source, along with the tools we created to maintain and update the SYCL version (since the CUDA source is still under active development).
During this process of migration and maintenance, we found that ``shuffle'' operations are not currently performance portable from NVIDIA to \Intel GPUs.
We developed a straightforward workaround to replace ``shuffles'' with local memory operations which we think could be generally useful for other developers.
This workaround could conceivably be added to SYCLomatic or SYCL compilers, and this replacement may not be necessary in future architecture generations.

With this project we have demonstrated the practical potential for writing performance-portable applications in SYCL, ultimately achieving a performance portability of 0.96 with near-zero code divergence.
Most of our experiments with architecture-specific tuning were written in pure SYCL and continued to be portable across architectures.
While a few of our kernel experiments involved vISA, even without resorting to inline assembly we were still able to achieve a performance portability of 0.91.
We continue in our efforts to apply optimizations identified in CRK-HACC's top hot spots to \emph{all} kernels to further improve performance portability.
We may also be able to achieve higher overall performance by selectively applying different optimization strategies to different kernels.


Our experiences here will inform our performance portability planning for CRK-HACC.
While it was manageable to develop and maintain several architectural variants of a few kernels for the older gravity-only versions of HACC, the task of developing and maintaining 20-30 kernels on 2-3 different architectures for CRK-HACC is considerably more difficult.
The development of an automated migration pipeline has been a critical piece of our current strategy of running production simulations on Frontier with CUDA/HIP while preparing to run production simulations on Aurora using SYCL.
Using a single programming model could significantly reduce total development and maintenance effort, even with a modest amount of code variation between architectures.

The SYCL version of CRK-HACC is an exciting proof-of-concept for using a single programming model across GPUs from Intel, NVIDIA, and AMD without sacrificing performance.
LLVM-based SYCL toolchains also use the native CUDA/ROCm runtimes, maintaining access to critical profiling tools.
Although the beta quality of the AMD plugin prevents us from using the SYCL version of CRK-HACC for ongoing simulations during early access to Frontier, this work demonstrates that access to production-quality SYCL implementations and profiling tools during the early stages of deployment on future supercomputers could enable real portable performance for production versions of high-performance codes.




\begin{acks}
This research was supported by the Exascale Computing Project (17-SC-20-SC), a collaborative effort of the U.S. Department of Energy Office of Science and the National Nuclear Security Administration.
This research used resources of the Argonne Leadership Computing Facility, which is a DOE Office of Science User Facility supported under Contract DE-AC02-06CH11357.
This research used resources of the Oak Ridge Leadership Computing Facility at the Oak Ridge National Laboratory, which is supported by the Office of Science of the U.S. Department of Energy under Contract No. DE-AC05-00OR22725.
\end{acks}

\section*{Disclaimers}

{\footnotesize
\noindent Performance varies by use, configuration and other factors. Learn more at~\hfill\xspace\linebreak \url{https://www.intel.com/performanceindex}. \smallskip

\noindent Performance results are based on testing as of dates shown in configurations and may not reflect all publicly available updates.  See Appendix~\ref{sec:reproducibility} for configuration details.  No product or component can be absolutely secure. Intel does not control or audit third-party data.  You should consult other sources to evaluate accuracy. \smallskip

\noindent \textcopyright~Intel Corporation. Intel, the Intel logo, and other Intel marks are trademarks of Intel Corporation or its subsidiaries. Other names and brands may be claimed as the property of others. Khronos\textsuperscript{\textregistered} is a registered trademark and SYCL\textsuperscript{\texttrademark} and SPIR\textsuperscript{\texttrademark} are trademarks of The Khronos Group Inc. \smallskip

\noindent No license (express or implied, by estoppel or otherwise) to any intellectual property rights is granted by this document, with the sole exception that code included in this document is licensed subject to the Zero-Clause BSD open source license (OBSD), \url{http://opensource.org/licenses/0BSD}.

\bibliographystyle{ACM-Reference-Format}
\bibliography{bibliography}
}

\appendix
\section{Configuration}\label{sec:reproducibility}
\normalsize

Results were collected by Argonne National Laboratory. \smallskip


\noindent The following flags were used by all SYCL runs:\hfill\xspace\linebreak \texttt{-DHACC\_CUDA\_POLY\_ORDER=5 -DHACC\_CUDA\_BLOCK\_SIZE=128\hfill\xspace\linebreak-O3 –std=c++17 -fsycl -ffast-math} 

\smallskip

\noindent \texttt{-fsycl-id-queries-fit-in-int} was used to convey that the number of work-items in each kernel fits in a signed 32-bit integer.

\subsection{Aurora}

\noindent \textbf{Date}: September 15, 2023

\noindent \textbf{CPU}: 2 socket Intel\textregistered~Xeon\textregistered~CPU Max 9470C, HT on, Turbo on

\noindent \textbf{OS}: SUSE Linux Enterprise Server 15 SP4

\noindent \textbf{Kernel}: 5.14.21-150400.24.55-default

\noindent \textbf{GPU}: Intel\textregistered~Data Center GPU Max 1550

\noindent \textbf{Driver}: 20230622.1-agama-devel-647.9

\noindent \textbf{SYCL Compiler}: Intel\textregistered~oneAPI DPC++/C++ Compiler 2023.2.0\linebreak(2023.x.0.20230510)

\noindent \textbf{SYCL Flags}: \texttt{-DHACC\_SYCL\_SG\_SIZE=16}

\subsection{Polaris}

\noindent \textbf{Date}: August 3, 2023

\noindent \textbf{CPU}: AMD EPYC 7543P 32-Core Processor, HT on, Turbo on

\noindent \textbf{OS}: SUSE Linux Enterprise Server 15 SP3

\noindent \textbf{Kernel}: 5.3.18-150300.59.115-default

\noindent \textbf{GPU}: NVIDIA A100-SXM4-40GB

\noindent \textbf{Driver}: 470.103.04, CUDA 11.4

\noindent \textbf{CUDA Compiler}: cuda\_11.8.r11.8/compiler.31833905\_0

\noindent \textbf{CUDA Flags}: \texttt{-Xcompiler -O3,-fopenmp,-g -gencode\hfill\xspace\linebreak~arch=compute\_80,code=sm\_80 --use\_fast\_math}

\noindent \textbf{SYCL Compiler}:
\url{https://github.com/intel/llvm},\hfill\xspace\linebreak cee07d3d26520fcdaab33dfaf7759b59210675d4

\noindent \textbf{SYCL Flags}: \texttt{-DHACC\_SYCL\_SG\_SIZE=32 -pthread -Wall\hfill\xspace\linebreak-sycl-std=2020 -fsycl-targets=nvptx64-nvidia-cuda\hfill\xspace\linebreak-Xsycl-target-backend --cuda-gpu-arch=sm\_80}

\subsection{Frontier}

\noindent \textbf{Date}: September 25, 2023

\noindent \textbf{CPU}: AMD EPYC 7A53 64-Core Processor, HT on, Turbo on

\noindent \textbf{OS}: SUSE Linux Enterprise Server 15 SP4

\noindent \textbf{Kernel}: 5.14.21-150400.24.46\_12.0.83-cray\_shasta\_c

\noindent \textbf{GPU}: AMD Instinct MI250X

\noindent \textbf{Driver}: HSA 1.1

\noindent \textbf{HIP Compiler}: roc-5.3.0 22362\hfill\xspace\linebreak3cf23f77f8208174a2ee7c616f4be23674d7b081

\noindent \textbf{HIP Flags}: \texttt{-fopenmp -O3 --amdgpu-target=gfx90a\hfill\xspace\linebreak-ffast-math}

\noindent \textbf{SYCL Compiler}: \url{https://github.com/intel/llvm},\hfill\xspace\linebreak 474461cb2e1de6f4e63a5d7d6edb6e93cb37b486

\noindent \textbf{SYCL Flags}: \texttt{-DHACC\_SYCL\_SG\_SIZE=64 -DNDEBUG\hfill\xspace\linebreak-sycl-std=2020 -fsycl-targets=amdgcn-amd-amdhsa\hfill\xspace\linebreak-Xsycl-target-backend=amdgcn-amd-amdhsa\hfill\xspace\linebreak--offload-arch=gfx90a}

\smallskip

\noindent Defining \texttt{-DNDEBUG} avoided a crash using \texttt{assert} in device kernels.

\end{document}